\documentclass[useAMS,usenatbib]{mn2e}
\usepackage{graphicx}

\title[Chromospheric self-heating in slowly rotating stars]{Self-heating and its possible relationship
to chromospheric heating in slowly rotating stars}

\author[Andria Rogava, Zaza Osmanov and Stefaan Poedts]{Andria Rogava$^{1,2,3}$\thanks{E-mail:
arogava@ictp.it} Zaza Osmanov$^{2}$\thanks{E-mail:
z.osmanov@astro-ge.org} and Stefaan Poedts$^{1}$\thanks{E-mail:
Stefaan.Poedts@wis.kuleuven.be}\\
$^{1}$Centre for Plasma Astrophysics, Katholieke Universiteit
Leuven, Celestijnenlaan 200B, bus 2400
B-3001, Belgium\\
$^{2}$Abdus Salam International Centre for Theoretical Physics,
Strada Costiera 11 34014 Trieste, Italy\\
$^{3}$Georgian National Astrophysical Observatory, Chavchavadze State University, Kazbegi 2a, Tbilisi, Georgia}
\begin{document}


\pagerange{\pageref{firstpage}--\pageref{lastpage}} \pubyear{2009}

\maketitle

\label{firstpage}

\begin{abstract}
The efficiency of nonmodal self-heating by acoustic wave
perturbations is examined. Considering different kinds of
kinematically complex velocity patterns we show that nonmodal
instabilities arising in these inhomogeneous flows may lead to
significant amplification of acoustic waves. Subsequently, the
presence of viscous dissipation damps these amplified waves and
causes the energy transfer back to the background flow in the form of
heat; viz. closes the ``self-heating'' cycle and contributes to the
net heating of the flow patterns and the chromospheric network as a
whole. The acoustic self-heating depends only on the presence of
kinematically complex flows and dissipation. It is argued that
together with other mechanisms of nonlinear nature the self-heating
\textit{may be} a probable additinal mechanism of nonmagnetic chromospheric
heating in the Sun and other solar-type stars with slow rotation and
extended convective regions.
\end{abstract}


\section{Introduction}

It is widely known that stars with outer convective zones, in terms
of their chromospheric line emission, exhibit twofold behavior.
Younger, rapidly rotating stars have the power of the chromospheric
emission higher by 2-3 orders of magnitude than slowly rotating,
older stars \citep{noy84}. It is believed that the emission in these
two classes of stars is related with different physical processes:
the low emission of slow rotators represents the effect of acoustic,
nonmagnetic heating, while the higher emission of fast rotators is
caused by effects of magnetic activity \citep{schr89}. In the last
decade it became increasingly clear that various slowly rotating,
solar-type stars, with well-developed convection zones, have similar
chromospheric emission. The available observational evidence
suggests \citep{g04} that this emission is caused, at least
partially, by the upward propagation and dissipation of sound
(acoustic) waves.

On the other hand, statistical analysis of existing ultraviolet (UV)
emission data of solar-type stars across the H-R diagram, has shown
the presence of a characteristic, well-pronounced ``basal" level of
\emph{chromospheric heating} \citep{jc98}. It is commonly assumed
that the heating may be caused by the nonlinear acoustic wave
shock-heating mechanism, but the dynamics of this phenomenon is not
completely understood. Moreover, the direct examination of Goddard
High-Resolution Spectrograph data of several evolved stars, having
similar ``basal" levels of chromospheric activity, failed to detect
a definite evidence in favor of the presence of acoustic waves,
presumably a key component of the heating mechanism. Instead, in a
number of cases, a counter-intuitive behavior, reminiscent of the
\emph{solar transition region} was detected, suggesting rather
\emph{a magnetic heating mechanism} for these stars. This
controversy led to the tentative conclusion that upward-propagating
shock waves do not necessarily dominate the observed radiative
losses from chromospheres of stars exhibiting typical ``basal"
behaviour. More generally, the nonmagnetic nature of the basal
components of convective, solar-type stars has been called in
question.

In solar physics the problem of the chromospheric heating has been
discussed ever since it was found out that the chromosphere is
hotter than the photosphere. Originally it was proposed \citep{bier,
schw} that the solar chromosphere is heated via the dissipation of
acoustic waves generated by the overshooting convection zone
turbulence. A theoretical model, determining the spectrum and power
of these waves, had been developed on the basis of the
Lighthill-Stein theory \citep{light52, stein67} of the convective
zone acoustic wave emission. It was suggested that the dissipation
of turbulently created acoustic waves may explain the observed
chromospheric heating in the Sun. In this model it was supposed
\citep{kalk07} that the heating takes place in the granules.
Alternatively, it may also occur in the lanes between granules
\citep{lites}, where the acoustic monopole emission may induce the
chromospheric oscillations. At any rate both these scenarios were
based on the assumption, confirmed by various observations of the
solar chromosphere, that the acoustic mechanism of heating may be
important in regions where the magnetic field is of a little
significance both dynamically and energetically.

However, similarly to the situation with other solar-type stars,
this explanation of the chromospheric heating has difficulties also
for the Sun. The analysis of solar data from the
\emph{SUMER}\emph{(Solar Ultraviolet Measurements of Emitted
Radiation)} instrument on SOHO, analogously to the above-mentioned
data from other slowly rotating, solar-type stars, revealed serious
problems with the nonlinear acoustic-shock-wave interpretation and
has underlined the necessity for the further investigation of this
issue. Observations with \emph{TRACE} \emph{(Transition Region And
Coronal Explorer)} NASA space telescope - revealed a very
substantial (about $90\%$!) deficit in the energy flux required to
heat the nonmagnetic, quiet solar chromosphere
\citep{fos1,fos2,fos3}. Later some light on this puzzle was shed by
numerical simulations with a three-dimensional hydrodynamic code. It
was argued \citep{wed07,cun07} that this puzzling discrepancy may
arise due to the limited spatial resolution of the \emph{TRACE},
failing to detect most of the acoustic waves traveling in quiet-Sun
inter-network regions of the chromosphere. In particular,
\citet{cun07} discussed the problems of the limited sensitivity of
TRACE while studying the three-dimensional solar chromospheric
topology; they reconsidered the acoustic chromospheric wave energy
flux and related this problem with the heating and emission of
chromospheric basal flux for $\tau$ Ceti type stars. It was
concluded that, contrary to claims by Fossum \& Carlsson,
high-frequency acoustic waves are sufficient to heat the nonmagnetic
solar chromosphere.

Recently various alternative scenarios of the chromospheric heating
for solar-type, cool stars with partially ionized chromospheres have
been proposed. It was argued that the Farley-Buneman \citep{f63}
plasma instability\footnote{This instability is responsible for the
heating of the E layer of the Earth ionosphere, which in a certain
sense is physically similar to the solar chromosphere.} may provide
the mechanism for conversion of the energy of convective motions of
neutral atoms into the chromospheric heating \citep{f05, f08}.
However \citet{g08}, taking into account the finite magnetization of
the ions and Coulomb collisions, re-evaluated the instability for
the solar chromosphere and argued that it could hardly be fully
responsible for the observed level of quasi-steady heating.

Therefore, the problem of the chromospheric heating remains an
actual and unsolved problem in the physics of slowly rotating,
solar-type stars and the Sun. Putting the problem in plain words, it
is observationally obvious that the chromospheres of stars with
extended convective zones are hotter than their photospheres. This
is true for both younger, rapidly rotating stars with magnetically
active chromospheres and for older stars, with slower rotation rates
and hydrodynamic, nonmagnetic chromospheres. \citet{kalk07} has
recently considered both possible mechanisms - the magnetic and the
nonmagnetic (acoustic) ones - that might be responsible for the
solar chromospheric heating. He argued \citep{kalk08a} that the
emission of the nonmagnetic chromosphere exhibits some
characteristic features of acoustic waves and no signatures of
magnetic waves.

Still, in general, the problem \emph{is }complicated. From the
existing observational, theoretical and numerical studies of the
``chromospheric heating problem" for cool, slowly rotating,
solar-type stars, it turns out that the problem doesn't have a
single, broadly accepted solution. Rather, it seems likely that
there may exist several different channels (mechanisms) of the
chromospheric heating. It is a challenging task to specify these
mechanisms and to provide at least a qualitative description of
their relative importance. It seems evident that at any rate the
chromospheric heating by acoustic waves takes place. The question is
whether nonlinear heating by means of acoustic shock-waves is the
only, or dominant, hydrodynamic channel of the heating or there are
other acoustic processes of the energy transfer from waves to the
medium that could lead to the chromospheric heating.

The role of the chromospheric \emph{flows} and their kinematic
complexity is possibly one of those important issues that has to be
tested in the chromospheric heating context. It is well-known that
the structure of the solar chromosphere is approximately spherically
symmetric for about the first 2000 km. Higher layers show complex
and variable fine structure, consisting of a wide variety of
\emph{flow patterns} - spicules, fibrils, surges
\citep{bl74,a86,zir88}. Recent observations by \emph{Hinode}
strengthened this picture, showing that the chromosphere often hosts
giant $H\alpha$ jets \cite{kos07,nish08}. It is now widely believed
that the solar chromosphere, both in its active and quiet
phases\footnote{The same is true for similar layers of the
atmospheres of stars with extended convective zones.} is highly
dynamic and highly inhomogeneous consisting of numerous flows
(``velocity patterns") with different spatial/temporal time-scales
and different geometries. These flows are closely related to the
\emph{chromospheric network } - a web-like structure seen in the
emissions of the red line of hydrogen ($H\alpha$) and the
ultraviolet line of calcium (Ca~II~K). The network is formed and
maintained due to the presence of bundles of magnetic field lines
ceaselessly shuffled and reshuffled by the fluid motions.

In particular, common ingredients of the solar chromosphere are
\emph{spicules} (small, jet-like ejections commonly observed
throughout the chromospheric network), \emph{macro-spicules} and
\emph{solar tornadoes} \citep{pm98,rog00}. These flows are
characterized by spatially inhomogeneous velocity fields: shear
flows with a nontrivial geometry and a poorly known but considerable
kinematic complexity. It is well-known that collective phenomena in
shear flows are strongly affected by so-called \emph{nonmodal
processes}, related with the non-self-adjointness of the operator
describing the linear dynamics of perturbations \citep{tref}. These
phenomena are especially robust and versatile in flows with
nontrivial kinematics and geometries. In particular, it has been
shown \citep{mahand} that under favourable conditions a
hydrodynamic, linear system may exhibit strongly pronounced unstable
regimes, where due to the efficient amplification of waves a large
part of the mean flow kinetic energy is pumped into the unstable
modes fostered by the flows. On the other hand, it was demonstrated
that if viscous dissipation is present, the large amplitude,
nonmodally amplified waves undergo viscous dissipation, leading to
the conversion of the wave energy into thermal energy and
corresponding \emph{``self-heating"} \citep{androSH} of the
``parent" flows.

The physics of the self-heating in a differentially moving medium
(shear flow) is quite generic and comprises the following three
steps: (a) The waves, originally, are excited spontaneously within
the shear flow; (b) They undergo nonmodal amplification, they grow,
extracting a part of the mean flow's kinetic energy. These processes
are quite efficient in neutral fluids \citep{bf92,ch97} and both in
electrostatic \citep{rmb97} and magnetized plasmas \citep{rog00}.
(c) Nonmodally amplified large-amplitude waves undergo, as the final
stage of their evolution, viscous decay: they get damped and give
their energy back to the ``parent" flow in the form of heat.

Evidently the crucial element of this three-step process is the
second one: nonmodal amplification of waves by the flow. The first
and the last ones - spontaneous excitation of waves and their
viscous damping - are routinely taking place in all kinds of
continuous media. The possibility of the nonmodal amplification of
collective modes in shear flows, however, creates a chance for the
waves to amplify \emph{before} they get damped! The energy needed
for the amplification is drawn from the ``reservoir" of the shear
flow kinetic energy \citep{bf92}. Eventual damping of these
larger-amplitude waves and conversion of their excessive energy into
the heat leads to the overall heating of the medium. Since this
process converts, via the agency of nonmodally amplified waves, the
kinetic energy of the flow into its thermal energy the process can
be called self-heating.

Self-heating was originally described for acoustic waves in a
hydrodynamic shear flow \citep{androSH}, but later it was found
\citep{lcl06,spp06} that self-heating occurs in magnetohydrodynamic
(MHD) shear flows too. The particularly robust nature of the
nonmodal amplification in kinematically complex flows
\citep{mahand,andro} suggested that self-heating could be especially
well pronounced in flows with complicated geometry and kinematics.

Since the solar chromosphere reveals the abundant presence of
various nontrivial velocity patterns, and there is no reason to
believe that chromospheric zones of other solar-type stars are any
different it is reasonable to investigate the above mentioned
mechanism in the context of the chromospheric heating for the slowly
rotating, solar-type stars. In the present paper we consider the
nonmodal self-heating mechanism in hydrodynamic, kinematically
complex flows and study the energy dissipation rate of acoustic
waves in the context of the possible relevance of this process to
chromospheric flows.

The paper is arranged in the following way: In section~II, we derive
the equations governing the linear evolution of perturbations of the
hydrodynamic system and consider the self-heating occurring within a
kinematically complex flow. In section~III, we apply this model to
the chromospheric flows and solve the set of equations numerically
for different parameter regimes. In section~IV, we discuss the
obtained results and outline the directions of further, more
detailed, quantitative and applicative studies.

\section[]{Main Consideration}

In order to show the efficiency of the self-heating for acoustic
waves generated, nonmodally amplified and viscously dissipated in
kinematically complex shear flows, we consider the following
standard hydrodynamic set, consisting of the equations of mass
conservation:
\begin{equation}
\label{conn} D_t\rho + \rho \nabla \cdot {\bf  V}= 0,
\end{equation}
the momentum conservation:
\begin{equation}
\label{euler} D_t{\bf V} = -\frac{1}{\rho}{\bf \nabla P} +
\nu\Delta{\bf V},
\end{equation}
and the polytropic equation of state:
\begin{equation}
\label{eqstate} P=C{\rho}^{1+1/n}.
\end{equation}

In these equations, $D_t \equiv \partial_t + ({\bf V} \cdot \nabla)$
denotes the convective derivative, $\rho$ is the density, ${\bf V}$
is the velocity, $\nu$ is the coefficient of kinematic
viscosity and $n$ is the polytropic index.

The instantaneous values of the physical variables are expressed as
sums of their equilibrium and perturbational components:
\begin{equation}
\label{dec1} \rho \equiv \rho_0 + \rho',
\end{equation}
\begin{equation}
\label{dec2} {\bf V} \equiv {\bf V_0} + {\bf V'},
\end{equation}
\begin{equation}
\label{dec3} P \equiv P_0 + P'.
\end{equation}
In the present study, for the sake of simplicity, the density of the
fluid in the unperturbed state is assumed to be homogeneous ($\rho_0
= \it{const}$).

Applying the standard procedure of linearization to
Eqs.~(\ref{conn},\ref{euler} and \ref{eqstate}) in terms of
Eqs.~(\ref{dec1},\ref{dec2} and \ref{dec3}) one derives the
following set of equations for the perturbed quantities:
\begin{equation}\label{con}
\mathcal{D}_t\rho' + \rho_0 (\nabla \cdot {\bf V'})= 0,
\end{equation}
\begin{equation}\label{eul}
\mathcal{D}_t{\bf V'} + ({\bf V'} \cdot \nabla){\bf V_0} =
-\frac{C_s}{\rho_0}{\bf \nabla \rho'} + \nu\Delta{\bf
V'},\end{equation}
where $\mathcal{D}_t \equiv \partial_t + ({\bf V_0}\cdot \nabla)$
and $C_s^2 = dP_0/d\rho_0$.

Since there is no first-hand information about the velocity fields
$\bf V_0$ of chromospheric flows and it is impossible to specify
`typical' or `standard' flow patterns we do not limit this study with a
narrowly chosen model for a prototype ``parent" flow. Instead we
will follow the approach originally developed and used in
\citet{mahand} which applies to a broad range of possible flow
patterns. Doing so we expand the velocity in a Taylor series in the
vicinity of a point $A(x_0,y_0,z_0)$. Preserving only linear terms
we write:
\begin{equation}\label{velexpand}
{\bf V}={\bf V}(A)+\sum_{i=1}^3\frac{\partial{\bf V}(A)}{\partial
x_i}(x_i-x_{i0}),\end{equation}
where $i=1,2,3$ and $x_i=(x,y,z)$.

In \citet{mahand} it was shown that the initial set of
linearized partial differential equations (\ref{con},\ref{eul}) is
transformed to the set of \emph{ordinary} differential equations by means of the
usage of the following ansatz for
the system's physical variables\footnote{Note, that in hydrodynamics
similar method, designed for the study of {\it incompressible}
perturbations in flows with spatially uniform shearing rates, has
been introduced by \citet{lan84} and \citet{cc86}.}:
\begin{equation}\label{anzatz}
F(x,y,z,t)\equiv\hat{F}(t)e^{\phi_1-\phi_2},\end{equation}
\begin{equation}\label{fi1}
\phi_1\equiv\sum_{i=1}^3{K_i}(t)x_i,\end{equation}
\begin{equation}\label{fi2}
\phi_2\equiv\sum_{i=1}^3V_i(A)\int{K_i}(t)dt,\end{equation}
where $V_i(A)$'s are the unperturbed velocity components and $K_i(t)$'s
are the wave vector components. The latter satisfy the following set of
ordinary differential equations:
\begin{equation}\label{dk}
{\bf \partial_{t}K}+ {\bf S^T} \cdot {\bf K}=0,\end{equation}
where ${\bf S^T}$ is the transposed shear matrix with the shear
matrix ${\bf S}$ \citep{mahand} defined as:
\begin{equation}\label{S}
 {\bf S} = \left(\begin{array}{ccc} V_{x,x} & V_{x,y} & V_{x,z}  \\
V_{y,x} & V_{y,y} & V_{y,z}  \\ V_{z,x} & V_{z,y} & V_{z,z} \\
\end{array} \right ),\end{equation}
with $V_{i,k}\equiv\partial V_{i}/\partial x_k$

Therefore, the ansatz reduces the mathematical aspect of the problem
to the study of the \emph{initial value problem} for the amplitudes
$\hat{F}(t)$ of the spatial Fourier harmonics of the perturbations.

The background flow velocity comprises
stretching and rotation of the flow field lines in the $XOY$ plane
superimposed on a spatially inhomogeneous outflow along the $Z$ axis.
In terms of the shear matrix it is written as \citep{andro,chven}:
\begin{equation}\label{S1}
 {\bf S} = \left(\begin{array}{ccc} \Sigma & A_1 & 0  \\
A_2 & -\Sigma & 0  \\ C_1 & C_2 & 0 \\
\end{array} \right ).\end{equation}

Using the notation $f^{(1)}\equiv df/d\tau$ for temporal
derivatives, where $\tau\equiv tK_nC_s$ is a dimensionless time
variable, we can write down the equations for the perturbations,
i.e.\ Eqs.~(\ref{con},\ref{eul}), in the following, completely
dimensionless form:
\begin{equation}\label{cont}
d^{(1)}+{\bf k} \cdot {\bf u} = 0,\end{equation}
\begin{equation}\label{v}
{\bf u}^{(1)}+{\bf S}\cdot{\bf u} = {\bf k} d-\overline{\nu}k^2 {\bf
u},\end{equation}
\begin{equation}\label{k}
\textbf{k}^{(1)} + {\bf S^T} \cdot {\bf k} = 0, \end{equation}
where  the following dimensionless notations are used: $d \equiv
-i\rho'/\rho_0$, $\textbf{u} \equiv \textbf{V}'/C_s$, $\sigma \equiv
\Sigma/K_nC_s$, $a_{1,2} \equiv A_{1,2}/K_nC_s$, $R_{1,2} \equiv
C_{1,2}/K_nC_s$, $\textbf{k}\equiv \textbf{K}/K_n$ and
$\overline{\nu}\equiv\nu K_n/C_s$. Here $K_n$ is the value that will
be used for the normalization of the wave vector. In particular, for
the two-dimensional case we take  $K_n \equiv K_x(0)$, while for the
three-dimensional case - $K_n \equiv K_z$($=const$).

For the sake of the forthcoming analysis it is useful to define the
total energy of the perturbations $E_{tot}$ and its components: the
kinetic, $E_{kin}$, and the compressional, $E_c$, energies:

\begin{equation}\label{E}
E_{tot}=E_{kin}+E_c=\frac{\textbf{u}^2}{2}+\frac{d^2}{2}.
\end{equation}

It is straightforward to verify that:
\begin{equation}\label{de}
E^{(1)}= -{\bf u} ({\bf S} \cdot {\bf
u})-\overline{\nu}k^2u^2.\end{equation}

For the evaluation of the efficiency of the self-heating we will
also use the \emph{self-heating rate}, defined as \citep{androSH}:
\begin{equation}\label{psi}
\Psi(\tau) \equiv \frac{\overline{\nu}}{E(0)} \int_0^\tau
k^2(\tau')u^2(\tau')d\tau'.\end{equation}

\section{Results}

From previous studies it is known that self-heating is strong enough
only if the instabilities are robust and swift enough to
pump energy into the acoustic waves before they are dumped by the viscous
dissipation. At the other hand, the viscosity must be strong enough
to ensure the feedback mechanism, i.e.\ the transfer of the energy
acquired by the waves back to the flow via viscous damping. It is
reasonable to expect that strongly pronounced self-heating occurs
only when there is a certain balance between the instability and
viscous time-scales \citep{androSH,lcl06}. Neither in the case of
\emph{weak nonmodal amplification} (i.e.\ when the waves do not have
enough time to grow before the viscosity kicks in), nor in the case
of \emph{negligibly weak viscosity} (i.e.\ when the dissipation is
not strong enough to dump the waves and heat the matter), the
self-heating mechanism will not be efficient enough to ensure a
substantial heating of the flow. From Eqs.~(\ref{v}), by means of a
simple dimensional analysis, one can obtain  a value of the maximum
length-scale at which the viscous terms start becoming important:
$\lambda_{cr}\leq \beta\nu/C_s$ (where $\beta$ is a dimensionless
parameter). Below, we will see that an efficient self-heating can
only happen for relatively small $\lambda \le \lambda_{cr}$
length-scales.

In this section we intend to show how strong the self-heating could
be for flows of different geometry and kinematics. Instead of
restricting the consideration to a specific kind of chromospheric
flow pattern we try to show the potential significance of this
nonmodal process for various kinds of kinematically complex
hydrodynamic flows.

As a first example, we consider the simplest, two-dimensional case.
The Eqs.~(\ref{cont}-\ref{k}) then reduce to:
\begin{equation}\label{cont2}
d^{(1)}+k_xu_x+k_yu_y = 0,\end{equation}
\begin{equation}\label{vx2}
u_x^{(1)}+\sigma u_x+a_1u_y =
k_xd-\overline{\nu}k^2u_x,\end{equation}
\begin{equation}\label{vy2}
u_y^{(1)}+a_2 u_x-\sigma u_y =
k_yd-\overline{\nu}k^2u_y,\end{equation}
\begin{equation}\label{kx2}
k_x^{(1)}+\sigma k_x+a_2 k_y = 0, \end{equation}
\begin{equation}\label{ky2}
k_y^{(1)}+a_1 k_x-\sigma k_y = 0.\end{equation}
\begin{figure}
 \par\noindent
 {\begin{minipage}[t]{1.\linewidth}
 \includegraphics[width=\textwidth] {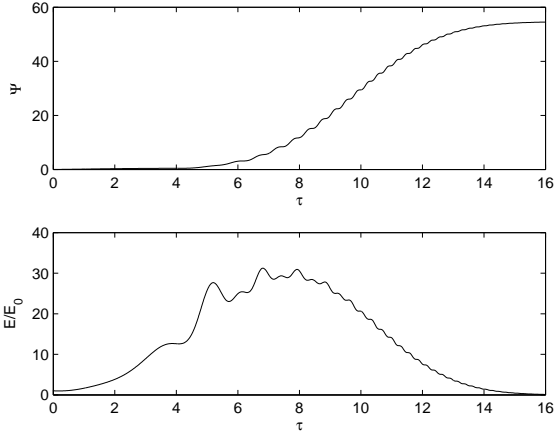}
 \end{minipage}
 }
 \caption[ ] {Temporal behaviour of the
self-heating rate and the normalized total energy of perturbation.
The set of parameters is: $a_1 = -1.3$, $\sigma=a_2 = 0$, $k_{x0} =
1$, $k_{y0} = -4$, $u_{x0} = 0.2$, $u_{y0} = 0$, $T = 6000K$,
$\lambda_x = 160km$, $n = 10^{11}cm^{-3}$.}\label{fig1}
 \end{figure}

Physical parameters of chromospheric flows in solar-type stars may
have broad ranges of variability. In the low solar chromosphere, for
instance, the flow velocity $V$, lies in the range $10-25\;$km/s ,
while the temperature, $T$, and the number density, $n$ are of the
order of $5000-6000\;$K and $10^{11}\;$cm$^{-3}$, respectively
\citep{sterl}. The length-scale, $R$, of chromospheric structures
may vary in the range $300-1500\;$km. One can easily show that for
these values the nonzero components of the shear matrix $\sim
V\lambda_x/( RC_s)$ [see Eq.~(\ref{S1})] for $\lambda_x\sim R$ are
of the order of unity\footnote{Here, $\lambda_x\sim 1/K_x(0)$}. For
the viscosity coefficient, we use the expression that is valid for a
weakly ionized plasma, which appears to be a good approximation for
the given range of temperatures \citep{visc}:
\begin{equation}\label{vis}
\nu = \frac{1}{3a_0^2n}\sqrt{\frac{k_BT}{m_p}},\end{equation}
where $a_0$ is the Bohr radius and $m_p$-the proton mass.

It is well-known that shear flows with a considerable kinematic
complexity are subject to various, nonmodal instabilities, sometimes
of a parametric nature \citep{mahand}. The nonmodal evolution of
perturbations strongly depends on the flow parameters and the wave
characteristics. For instance, it was shown \citep{andro,chven} that
if $\Gamma\equiv\sigma^2+a_1a_2> 0$, the system undergoes a linear
instability, while for $\Gamma\leq 0$ the flow remains stable and
exhibits an instability only for a narrow range of perturbation
parameters.
\begin{figure}
 \par\noindent
 {\begin{minipage}[t]{1.\linewidth}
 \includegraphics[width=\textwidth] {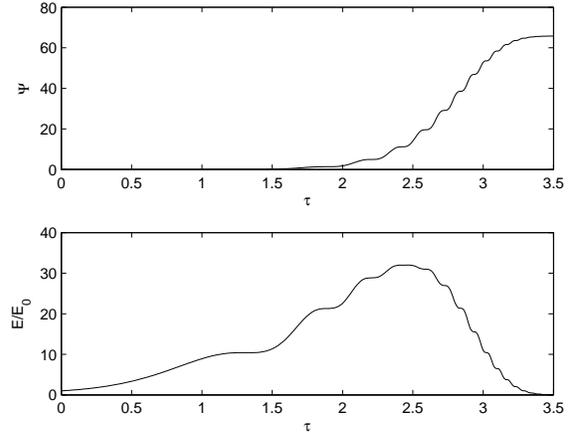}
 \end{minipage}
 }
 \caption[ ] {Temporal behaviour of the
self-heating rate and the normalized total energy of perturbation.
The set of parameters is: $a_1 = a_2 = -\sigma = 1$, $k_{x0} =
k_{y0} = 1$, $u_{x0} = 0.2$, $u_{y0} = 0$, $T = 6000K$, $\lambda_x =
200km$, $n = 10^{11}cm^{-3}$.}\label{fig2}
 \end{figure}

For the first example, we consider the case when the velocity shear
is one-dimensional. In Fig.~(\ref{fig1}) the time evolution of the
self-heating rate and the total energy of the acoustic perturbation
for the case of a parametric instability ($\Gamma\leq 0$) are
displayed. The set of parameters in this case was: $a_1 = -1.3$,
$\sigma=a_2 = 0$, $k_{x0} = 1$, $k_{y0} = -4$, $u_{x0} = 0.2$,
$u_{y0} = 0$, $T = 6000\;$K, $\lambda_x = 160\;$km, $n =
10^{11}\;$cm$^{-3}$. From these plots it is clear that initially the
unstable growth dominates over the viscous damping and the energy of
the perturbation steadily increases. Gradually, however, the
viscosity becomes more important and, as a result, the perturbation
energy reaches its maximum level and starts decreasing until the
perturbation completely disappears, giving back all the gained
energy to the flow. Note that in this case the asymptotic value of
the self-heating rate $\Psi_{\infty} \equiv \lim_{t \to
\infty}\Psi(t) \simeq 60$ is reached.

The second example shows the case [see Fig.~(\ref{fig2})] when the
wave number vector components evolve exponentially ($\Gamma>0$). The
set of parameters is: $a_1 = a_2 = -\sigma = 1$, $k_{x0} = k_{y0} =
1$, $u_{x0} = 0.2$, $u_{y0} = 0$, $T = 6000\;$K, $\lambda_x =
200\;$km, $n = 10^{11}\;$cm$^{-3}$. The qualitative behaviour of the
system is similar to the previous case: initially the perturbation
rapidly grows at the expense of the background flow energy. But due
to the exponential growth of the modulus of the wave vector the
perturbation length-scale rapidly decreases, leading to the
inevitable domination of the viscosity, which, in its turn,
eventually damps the perturbation, converting its excessive energy
into the thermal energy of the flow. The figure shows that the
asymptotic self-heating rate in this case is about
$\Psi_{\infty}\sim 70$.

Observational evidence suggests that the solar chromosphere is
populated by three-dimensional fluxes, viz.\ flow patterns with
nontrivial morphologies. An interesting class of these structures
are the \emph{swirling macrospicules}, also called the \emph{solar
tornadoes} \cite{pm98}, characterized by a helical motion of the
plasma. It is plausible to expect that similar, highly nontrivial
flow patterns may exist in chromospheric layers of other slowly
rotating, solar-type stars. These are particularly intricate
examples of flows with a very high degree of kinematic complexity!
They may host various kinds of strongly-pronounced nonmodal shear
instabilities \cite{andro,chven}. Therefore, in our analysis, it is
worthwhile to consider three-dimensional cases too and to check
whether self-heating processes are equally or more efficient in
them.

We consider the velocity configuration described by:
\begin{equation}\label{vel}
{\bf V(r)} \equiv [0, r\Omega(r), U(r)],\end{equation}
with $\Omega (r) = \mathcal{A}/r^{\alpha}$ and $U =\;$const. In this
case the shear matrix can be written in the following way
(\cite{andro}):
\begin{equation}\label{S1}
 {\bf S} = \left(\begin{array}{ccc} 0 & -\mathcal{A}/r^{\alpha} & 0  \\
\mathcal{A}\left(1-\alpha\right)/r^{\alpha} & 0 & 0  \\ R_1 & R_2 & 0 \\
\end{array} \right ).\end{equation}

\begin{figure}
 \par\noindent
 {\begin{minipage}[t]{1.\linewidth}
 \includegraphics[width=\textwidth] {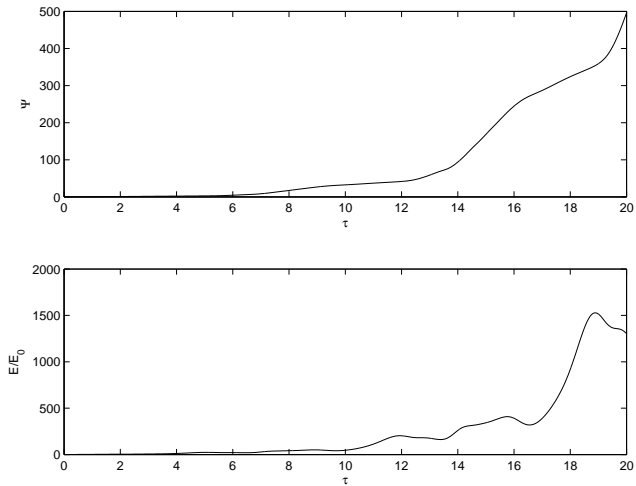}
 \end{minipage}
 }
 \caption[ ] {Temporal behaviour of the
self-heating rate and the normalized total energy of perturbation.
The set of parameters is: $\alpha=0.2$, $r = R/2$, $R_1 = R_2 = -1$,
$k_{z0} = -k_{x0} = 1$, $k_{y0} = -5$, $u_{x0} = 0.1$, $u_{y0} =
u_{z0} = 0$, $R = 500km$, $U = 20km/s$, $T = 6000K$, $\lambda_x =
200km$, $n = 10^{11}cm^{-3}$.}\label{fig3}
 \end{figure}

As an example, let us consider the case $\alpha=0.2$,
$V_{\varphi}(R) = U$, and $r = R/2$. Fig.~(\ref{fig3}) shows the
solution of Eqs.~(\ref{cont}-\ref{k}) for the following set of
parameters: $R_1 = R_2 = -1$, $k_{z0} = -k_{x0} = 1$, $k_{y0} = -5$,
$u_{x0} = 0.1$, $u_{y0} = u_{z0} = 0$, $R = 500km$, $U = 20km/s$, $T
= 6000K$, $\lambda_x = 200km$, $n = 10^{11}cm^{-3}$. From the plots
we see that the instability is so robust that the viscous
dissipation fails to affect the perturbation evolution in a
graphically \emph{visible} way. Nevertheless, despite the seemingly
insignificant contribution of viscous terms, their role in the
self-heating process is as much instrumental as in all previously
considered cases. The figure shows that the self-heating rate at the
end of the given time interval ($\tau = 20$) is quite high $\Psi\sim
500$.

\section{Discussion and Conclusions}

Slowly rotating, older stars with well-developed convection zones
(solar-type stars) feature similar chromospheric emission.
Observations indicate that the emission is related to the
dissipation of sound (acoustic) waves \citep{g04} propagating from
the photosphere to the chromosphere. The statistical analysis of the
UV emission data from these stars shows the presence of a
significant level of chromospheric heating \citep{jc98}. Commonly it
is assumed that the heating is due to the dissipation of nonlinear
acoustic waves. But the dynamics and the strength of the
shock-heating mechanism are not well-understood and the nonmagnetic
nature of chromospheric heating for solar-type stars has been
recently doubted. From Goddard High-Resolution Spectrograph data of
a number of solar-type stars no firm evidence was detected of the
presence of acoustic waves. Quite on the contrary, in some cases the
observational evidence seemed to be in favour of a magnetic
mechanism for chromospheric heating for these stars. On the basis of
this controversy it was surmised that upward-propagating shock waves
do not necessarily govern radiative losses from the chromospheres of
solar-type stars.

The situation is much the same in the context of solar physics. It
is commonly believed that the solar chromosphere, in its quiet
phase, in regions with negligibly weak magnetic field, has to be
heated by acoustic waves. However, the energy flux of these waves,
measured in the upper photosphere by TRACE, has been found to be
insufficient to explain the radiative emission from the
chromosphere. \citet{wed07} and \citet{cun07}, employing the
three-dimensional hydrodynamical model \citep{wed04}, have suggested
that the spatial resolution of the TRACE is insufficient to resolve
intensity fluctuations that occur on small spatial scales.

\citet{kalk07} has also considered this assumption, implying that
the spatial averaging by TRACE may serve as a qualitative
explanation for the observed acoustic flux deficit. However, he
found out that the standard hydrodynamical model was too
oversimplified in the treatment of chromospheric energy exchange to
provide a quantitative explanation of the suppression of the
acoustic fluctuations. The mentioned hydrodynamic test \citep{wed07}
was of a preliminary nature, because the hydrodynamic model on which
the test was based involved temperature fluctuations exceeding those
of the Sun. The shape of the acoustic spectrum observed with TRACE
seemed to support the theory of wave generation in the solar
convection zone even though the limited acoustic frequency range and
the low energy flux of the observations did not allow to make any definitive
conclusion \citep{kalk07,kalk08a}. That is why the conclusion was
that  \emph{``the heating mechanism of the chromosphere thus
remained an open question"} \citep{kalk08a}.

The main goal of our present study was to show that the nonmodal
self-heating by acoustic waves may serve as an additional source of
chromospheric heating. We demonstrated the possibility of the
efficient heating in chromospheric flows due to the presence of two
basic factors: the existence of a spatially inhomogeneous,
sufficiently complex velocity field (kinematically complex ``flow
patterns''), and the existence of viscous damping.

The self-heating is efficient when some necessary conditions upon the flow
characteristics, for instance the kinetic energy budget of the mean
flow, are met. The energy of the regular motion has to serve as an
ample energy reservoir for the growing waves. Therefore, it is
reasonable to consider the kinetic energy density of the flows
in chromospheric structures and to compare it with the
energy density of unstable modes.

Let us suppose that the characteristic velocity and the mass density
of a chromospheric structure (e.g., a macro-spicule, a solar
tornado, or a jet-like surge) are $V$ and $\rho$, respectively. The
kinetic energy density can be expressed by:
\begin{equation}\label{ekin}
E_{kin} \approx\frac{\rho V^2}{2}.\end{equation}

For the initial energy of the induced mode one gets [see
Eq.~(\ref{E})]:

\begin{equation}\label{ind}
E_{ind} =\frac{\rho C_s^2}{2}\left(u_0^2+d_0^2\right).\end{equation}

The energy converted into the form of heat, at the other hand, is of
the order:

\begin{equation}\label{eheat}
E_{heat} \approx \Psi\frac{\rho
C_s^2}{2}\left(u_0^2+d_0^2\right),\end{equation}
implying that each generated mode extracts a certain portion of the
mean flow kinetic energy:
\begin{equation}\label{eta}
\eta
\equiv\Psi\frac{C_s^2}{V^2}\left(u_0^2+d_0^2\right).\end{equation}

For the parameters used for Fig.~(\ref{fig1}) and Fig.~(\ref{fig2})
one can easily show that $\eta$ equals $\sim 0.5$ and for
Fig.~(\ref{fig3}), $\eta\sim 1$, leading to the conclusion that for
the cases shown in Fig.~(\ref{fig1}) and Fig.~(\ref{fig2}) about
$50\%$ of the chromospheric structures' kinetic energy has been
extracted and given back to the flow in the form of heat and almost
$\sim 100\%$ for the case presented in Fig.~(\ref{fig3}). It means
that this process is quite efficient for the heating of the ``parent"
flow. Apparently, if a large part of energy is extracted
from the flow patterns they could easily be destroyed by the
self-inflicted process of self-heating. This circumstance, in turn,
could contribute to the statistically average short lifetimes ($\sim
1-10\;$min) of the macrospicules. For example, the case shown in
Fig. (\ref{fig1}) is characterized by the typical timescale of the
self-heating, $t\sim \tau\lambda /C_s$ being of the order of $5min$.
On the other hand, during one complete cycle of the self heating,
almost $50\%$ of the flow kinetic energy will be extracted. For
Fig.~(\ref{fig2}) the same energy fraction is extracted from the
mean flow, but the heating timescale would be $\sim 2min$. Rather
different situation is shown in Fig.~(\ref{fig3}), where the
viscosity is not enough to saturate the instability and as is seen
from the plot, the perturbed energy is increased asymptotically.
Nevertheless, the presence of viscous damping is still significant,
because as we see from Fig.~(\ref{fig3}), the self-heating rate for
$\tau = 20$ equals $\sim 500$. On the other hand, we have already
mentioned that $\eta\sim 1$, thus the considered
mode can extract the mean flow kinetic energy in approximately
$7min$. As is seen from these figures, the more complex is the
kinematics of the flow, the more efficient is the self-heating
process.

The results of the present study, being of a qualitative nature,
indicate that nonmodal shear flow instabilities, coupled with the
presence of the viscous dissipation, may ensure the net
chromospheric heating by the agency of the flow patterns of
sufficiently complex kinematic structure. The presence of magnetic
fields cannot alter this qualitative picture. On the contrary, it is
known that in magnetized helical flows nonmodal shear instabilities
tend to become much more powerful \citep{andro,chven}; and it has
been shown that self-heating in the MHD may be quite efficient both
generally \citep{lcl06} and in the astrophysical context
\citep{spp06}. It makes us believe that if self-heating is able to
heat nonmagnetic part of the chromosphere of a solar-type star, it
will certainly do the same job in a magnetized chromosphere too. We
plan to study this issue in detail in the near future.

In the standard, Biermann-Schwarzschild, scenario of heating of the
solar chromosphere \citep{bier,schw} the velocity amplitude of
upward-propagating acoustic waves grows due to the exponential
decrease of the mass density with height; leading, in turn, to shock
formation and dissipation of the acoustic waves and heating of the
medium. The important feature of this mechanism is the exponential
dependence of the mass density on height, because for a medium with
$\rho=const$ this classic scenario does not work. Alternatively, the
nonmodal self-heating works when the density is homogeneous! On the other hand,
it crucially depends on the presence of the kinematically complex shear
flow. It is reasonable to expect that in realistic situations, viz.
in chromospheric structures where\emph{ both} density stratification
and flows are present, these two mechanisms will be complementary
and will serve as additional sources of the heating.

One of the goals of our future research will be also the study of
the propagation of acoustic gravity waves in gravitationally
stratified chromospheric flows in the presence of viscous
dissipation. Both linear and nonlinear regimes of acoustic-gravity
wave propagation in two- \citep{kalk94} and three-dimensional
\citep{bod00} hydrodynamics and in the presence of the temperature
inhomogeneity \citep{bod01} have been previously studied. It is
interesting to study the dynamics of these waves in the presence of
velocity shear and viscous damping and to check whether the nonmodal
self-heating influences these processes too.

In order to study the role of the shear-induced nonmodal instability
for the problem of the \emph{chromospheric heating} in slowly
rotating, solar-type stars we considered the model of a simple,
quite general velocity pattern and wrote down hydrodynamic equations
governing the evolution of acoustic perturbations within this flow.
We have solved these equations for several, representative examples
of flow patterns and found out that the instability and the viscous
damping might lead to quite efficient self-heating of the ``parent"
flows. Analyzing the kinetic energy budget of a typical
chromospheric macro-spicule, we found out that the self-heating
might extract a significant part of kinetic energy from the velocity
pattern in time-scales of the same order as the lifetimes of the
corresponding structures observed in the solar chromosphere.
Therefore, it is quite logical to believe that self-heating of
chromospheric flow patterns contributes to the overall heating of
the chromosphere by converting a considerable part of their kinetic
energy into heat and ultimately destroying these flow patterns.

\section*{Acknowledgments}

The research of AR and ZO was supported by the Georgian National
Science Foundation grant GNSF/ST07/4-193. AR is grateful to
Katholieke Universiteit Leuven for partial support by the Senior
Research Fellow award. These results were obtained in the framework
of the projects GOA/2009-009 (K.U.Leuven), G.0304.07
(FWO-Vlaanderen) and C~90205 (ESA Prodex 9). Financial support by
the European Commission through the SOLAIRE Network
(MTRN-CT-2006-035484) is gratefully acknowledged.

\bsp

\label{lastpage}


\begin{thebibliography}{99}
\bibitem[\protect\citeauthoryear{Athay}{1986}]{a86} Athay R.G., 1986, in Physics of the Sun,
Vol. II, ed. P. A. Sturrock (Dordrecht: Reidel), 51
\bibitem[\protect\citeauthoryear{Bierman}{1946}]{bier} Biermann L., 1946, Naturwissenschaften, 33, 118
\bibitem[\protect\citeauthoryear{Bodo et al.}{2000}]{bod00} Bodo G., Kalkofen W., Massaglia S., Rossi
P., 2000, A\&A, 354, 296
\bibitem[\protect\citeauthoryear{Bodo et al.}{2001}]{bod01} Bodo G., Kalkofen W., Massaglia S., Rossi
P., 2001, A\&A, 370, 1088
\bibitem[\protect\citeauthoryear{Bray \& Loughhead}{1974}]{bl74} Bray R.J., Loughhead,
R.E., 1974, The Solar Chromosphere (London: Chapman and Hall)
\bibitem[\protect\citeauthoryear{Butler \& Farrell}{1992}]{bf92} Butler K.M.,
Farrell B. F., 1992, Phys. Fluids A, 4,
1637
\bibitem[\protect\citeauthoryear{Chagelishvili et
al.}{1997}]{ch97} Chagelishvili G.D., Khujadze G.R., Lominadze J.G.,
Rogava A.D., 1997, Phys. Fluids, 7, 1955
\bibitem[\protect\citeauthoryear{Craik \& Criminale}{1986}]{cc86} Craik A.D.D., Criminale W.O., 1986, Proc.R.Soc.Lond.
A, 406, 13

\bibitem[\protect\citeauthoryear{Cuntz et al.}{2007}]{cun07} Cuntz M., Rammacher W.,
Musielak Z.E., 2007, ApJ, 657, L57
\bibitem[\protect\citeauthoryear{Farley}{1963}]{f63} Farley D.T.
Jr., 1963, JGR, 68, 6083
\bibitem[\protect\citeauthoryear{Fontenla}{2005}]{f05} Fontenla
J.M., 2005, A\&A, 442, 1099
\bibitem[\protect\citeauthoryear{Fontenla et al.}{2008}]{f08}
Fontenla J.M., Peterson W.K., Harder J., 2008, A\&A, 480, 839
\bibitem[\protect\citeauthoryear{Fossum \& Carlsson}{2005a}]{fos1}
Fossum A., Carlsson M., 2005a, ApJ, 625, 556
\bibitem[\protect\citeauthoryear{Fossum \& Carlsson}{2005b}]{fos2} Fossum A., Carlsson M., 2005b, Nature, 435, 919
\bibitem[\protect\citeauthoryear{Fossum \& Carlsson}{2006}]{fos3} Fossum A., Carlsson M., 2006, ApJ, 646, 579
\bibitem[\protect\citeauthoryear{Gogoberidze et al.}{2008}]{g08} Gogoberidze G., Voitenko Yu.,
Poedts S., Goossens M., 2009, arXiv:0902.4426
\bibitem[\protect\citeauthoryear{Goodman}{2000}]{visc} Goodman M.L., 2000, ApJ, 533, 501
\bibitem[\protect\citeauthoryear{Goodman}{2004}]{g04} Goodman M.L., 2004, A\&A, 424, 691
\bibitem[\protect\citeauthoryear{Kalkofen et al.}{1994}]{kalk94} Kalkofen W., Rossi P., Bodo G., Massaglia,
S., 1994, A\&A, 284, 976
\bibitem[\protect\citeauthoryear{Kalkofen}{2007}]{kalk07} Kalkofen W., 2007, ApJ, 671, 2154
\bibitem[\protect\citeauthoryear{Kalkofen}{2008}]{kalk08a} Kalkofen W., 2008b, J. Astrophys. Astr., 2008, 29,
163




\bibitem[\protect\citeauthoryear{Kosugi et al.}{2007}]{kos07} Kosugi T. et al., 2007, Sol. Phys.,
243, 3
\bibitem[\protect\citeauthoryear{Judge \& Carpenter}{1998}]{jc98} Judge P.G., Carpenter K.G., 1998, ApJ, 494,
828
\bibitem[\protect\citeauthoryear{Lagnado et al.}{1984}]{lan84} Lagnado R.R., Phan-Thien N., Leal, L.G., 1984, Phys.
Fluids, 27, 1094
\bibitem[\protect\citeauthoryear{Li et al.}{2006}]{lcl06} Li J.W., Chen Y., Li Z.Y., 2006, Phys. Plasmas, 13, 042101
\bibitem[\protect\citeauthoryear{Lighthill}{1952}]{light52} Lighthill M.J., 1952, Proc. R.
Soc. London, A, 211, 564
\bibitem[\protect\citeauthoryear{Lites et al.}{1993}]{lites} Lites B.W., Rutten R.J., Kalkofen W., 1993, ApJ, 414, 345
\bibitem[\protect\citeauthoryear{Mahajan \& Rogava}{1999}]{mahand} Mahajan S.M., Rogava A.D., 1999, ApJ,
518, 814
\bibitem[\protect\citeauthoryear{Nishizuka et al.}{2008}]{nish08} Nishizuka N., 2008, ApJ,
683, L83
\bibitem[\protect\citeauthoryear{Noyes et al.}{1984}]{noy84} Noyes R.E., Hartmann L.W.,
Baliunas S.L., Duncan, D.K., Vaughan, A.H., 1984, ApJ, 279, 763
\bibitem[\protect\citeauthoryear{Pike \& Mason}{1998}]{pm98} Pike C.D., Mason
H.E., 1998, Sol. Phys., 182, 333
\bibitem[\protect\citeauthoryear{Rogava et al.}{1997}]{rmb97} Rogava A.D., Mahajan S.M. Berezhiani V.I.,,
1997, Phys. Plasmas, 12, 4201
\bibitem[\protect\citeauthoryear{Rogava et al.}{2000}]{rog00} Rogava A. D., Poedts S.,
Mahajan S.M., 2000, A\&A, 354, 749
\bibitem[\protect\citeauthoryear{Rogava et al.}{2003a}]{andro} Rogava A.D., Mahajan S.M., Bodo G., Massaglia S., 2003, A\&A, 399,
421
\bibitem[\protect\citeauthoryear{Rogava et al.}{2003b}]{chven} Rogava A.D., Bodo G., Massaglia S., Osmanov, Z., 2003, A\&A, 408,
401
\bibitem[\protect\citeauthoryear{Rogava}{2004}]{androSH} Rogava A. D., 2004, Ap. Space Sci., 293, 189
\bibitem[\protect\citeauthoryear{Schrijver et al.}{1989}]{schr89} Schrijver C.J., Dobson A.K., Radick R.R., 1989, ApJ, 341, 1035
\bibitem[\protect\citeauthoryear{Schwarz\-schild}{1948}]{schw} Schwarzschild M., 1948, ApJ, 107, 1
\bibitem[\protect\citeauthoryear{Shergelashvili et al.}{2006}]{spp06} Shergelashvili B.M., Poedts S., Pataraya A.D.,
2006, ApJ, 642, L73
\bibitem[\protect\citeauthoryear{Stein}{1967}]{stein67} Stein R.F., 1967, Sol. Phys., 2, 385
\bibitem[\protect\citeauthoryear{Sterling}{2000}]{sterl} Sterling A. C., 2000, Sol. Phys., 196, 79
\bibitem[\protect\citeauthoryear{Trefethen et al.}{1993}]{tref} Trefethen L.N., Trefethen A.E., Reddy S.C. Driscoll T.A., 1993,
Sience, 261, 578
\bibitem[\protect\citeauthoryear{Wedemeyer-B\"{o}hm et al.}{2004}]{wed04} Wedemeyer-B\"{o}hm S. et al., 2004, A\&A,
414, 1121
\bibitem[\protect\citeauthoryear{Wedemeyer-B\"{o}hm et al.}{2007}]{wed07} Wedemeyer-B\"{o}hm S.,
Steiner O., Bruls J., Rammacher, W., 2007, in ASP Conf. Ser. 368,
The Physics of Chromospheric Plasmas, ed. P. Heinzel, I. Dorotovic,
\& R.J. Rutten (San Francisco: ASP), 93
\bibitem[\protect\citeauthoryear{Zirin}{1988}]{zir88} Zirin H., 1988, Astrophysics of the Sun
(Cambridge: Cambridge Univ. Press)
\end{thebibliography}
\end{document}